\documentclass[prl,aps,twocolumn,notitlepage,superscriptaddress,showpacs,nofootinbib]{revtex4-2}
\usepackage{amsmath}
\usepackage{natbib}
\usepackage{rotating}
\usepackage{extarrows}
\usepackage{graphicx}
\usepackage{dcolumn}
\usepackage{bm}
\usepackage[all]{xy}
\usepackage{indentfirst}
\usepackage{xcolor, soul}
\usepackage{soul}
\usepackage{ulem}
\usepackage{amsmath}
\usepackage{bbm}
\usepackage{color}
\usepackage{multirow}
\usepackage{mathrsfs}
\usepackage{euscript}
\usepackage{amssymb}
\usepackage[colorlinks]{hyperref}
\hypersetup{
	colorlinks = true,
	urlcolor = {blue},
	citecolor = {blue},
	linkcolor= {blue}
}

\newcommand{\ket}[1]{|#1\rangle}

\newcommand{\hb}{{b^{\prime}}}
\newcommand{\hc}{{c^{\prime}}}
\newcommand{\hz}{{{z}^{\prime}}}

\newcommand{\va}{\mathbf{a}}
\newcommand{\vx}{\mathbf{x}}

\newcommand{\dt}{\diamondsuit}

\newcommand{\oI}{\overline{I}}

\newcommand{\hsC}{{\textsf{C}^{\prime}}}

\newcommand{\hB}{{{B}^{\prime}}}

\newcommand{\sA}{\textsf{A}}
\newcommand{\sB}{\textsf{B}}
\newcommand{\sC}{\textsf{C}}

\newcommand{\cH}{\mathcal{H}}

\newtheorem{theorem}{Theorem}

\begin{document}
\title{Testing Genuine Multipartite Nonlocality via an Inflated Network with Multi-copy Entangled States}

\author{Qian-Xi Zhang}
\altaffiliation{These authors contributed equally to this work.}
\affiliation{International Quantum Academy, Shenzhen, 518048, China}
\affiliation{Southern University of Science and Technology, Shenzhen, 518055, China}

\author{Ming-Xing Luo}
\altaffiliation{These authors contributed equally to this work.}
\affiliation{School of Information Science and Technology, Southwest Jiaotong University, Chengdu 610031, China}
\affiliation{Hefei National Laboratory, Hefei, Anhui 230088, China}

\author{Ya-Li Mao}
\email{maoyl@nankai.edu.cn}
\affiliation{School of Physics, Nankai University, Tianjin, 300071, China}

\author{Hu Chen}
\affiliation{International Quantum Academy, Shenzhen, 518048, China}
\affiliation{Southern University of Science and Technology, Shenzhen, 518055, China}

\author{Yu-Hang Yao}
\affiliation{International Quantum Academy, Shenzhen, 518048, China}
\affiliation{Southern University of Science and Technology, Shenzhen, 518055, China}

\author{Zhi-Lian Liu}
\affiliation{International Quantum Academy, Shenzhen, 518048, China}
\affiliation{Southern University of Science and Technology, Shenzhen, 518055, China}

\author{Shao-Ming Fei}
\affiliation{School of Mathematical Sciences, Capital Normal University, Beijing 100048, China}
\affiliation{Max-Planck-Institute for Mathematics in the Sciences, 04103 Leipzig, Germany}

\author{Xue Yang}
\email{yx12290552@swjtu.edu.cn}
\affiliation{School of Information Science and Technology, Southwest Jiaotong University, Chengdu 610031, China}

\author{Zheng-Da Li}
\email{lizhengda@iqasz.cn}
\affiliation{International Quantum Academy, Shenzhen, 518048, China}

\begin{abstract}
Understanding the nonlocality of multipartite quantum systems provides valuable insights into their behaviors and  potential applications. In this Letter, assuming a quantum network inflated with multiple copies of genuine multipartite entangled states, we propose a novel noise-robust approach to test the genuine multipartite nonlocality inherent in each copy under Svetlichny's biseparable model. This extends Gisin's Theorem to an arbitrary number of parties, establishing the equivalence among genuine multipartite nonlocality, genuine multipartite steering, and genuine multipartite entanglement for all multipartite pure states under multiple copies assumption. In the experiment, we employ a hybrid photonic quantum network to verify the genuine tripartite nonlocality of generalized Greenberger-Horne-Zeilinger (GHZ) and W states beyond previously explored parameter regimes. This work not only offers a unified robust method on exploring multipartite quantum correlations, but also opens a new avenue for studying genuine multipartite nonlocality through network-distributed multi-copy quantum states and different network topologies.

\end{abstract}
\maketitle

Bell's theorem demonstrates that entangled quantum systems can exhibit correlations that violate certain constraints satisfied by any local realistic theory \cite{Bell}, providing a new experimentally testable method for resolving the question raised by the Einstein-Podolsky-Rosen (EPR) argument \cite{EPR}. This was later generalized by Gisin \cite{GP} who showed all pure entangled states can violate the Clauser-Horne-Shimony-Holt (CHSH) inequality \cite{CHSH}. However, this result does not hold for generalized noisy states \cite{Werner, HHH}. 

A form of nonlocality for multipartite systems stronger than standard Bell nonlocality is genuine multipartite nonlocality (GMN), which cannot be reproduced by classical mixtures of bipartite separable correlations \cite{Sy, Mermin, ardehali1992bell, belinskiui1993interference, gisin1998bell, RMP}. The verification of GMN typically relies on specialized Bell-type inequalities, such as the Svetlichny inequality \cite{Sy} or Mermin-Ardehali-Belinskii-Klyshko (MABK) inequalities \cite{Mermin, ardehali1992bell, belinskiui1993interference}. Recently, numerous experiments have successfully demonstrated GMN across a range of physical platforms \cite{pan2000experimental, PhysRevLett.91.180401, lavoie2009experimental, PhysRevLett.112.100403, hamel2014direct, Zhang2016, Wang2016, Huang12022}. 

Gisin's concept has also been extended to characterize many-body systems \cite{PR, Cola, Fei, Hardy, Chou, Yu, Chen, curchod2019versatile, stachura2024single}. Interestingly, Scarani and Gisin showed that generalized Greenberger-Horne-Zeilinger (GHZ) states do not violate the MABK inequalities \cite{SG}, and similar results hold for (odd) $n$-particle GHZ states \cite{Zuko} or general genuine multipartite entanglement (GME) \cite{Gal, Ban, Waldemar2019, Designolle2017, ZD, GN, AN, ZY, CL, RD}, depending on specific measurement settings employed \cite{NJ, JN, Chen2014, VL}. This implies the nonequivalence of GMN and GME even for isolated systems. However, a recent result shows the joint system of $n-1$ copies of a GME state exhibits GMN \cite{Contreras2021,Contreras2022}. This approach does not verify the GMN of a single copy and lacks noise robustness, and therefore cannot be applied to noisy quantum states and experiments.

\begin{figure}
\begin{center}
\includegraphics[width=0.46\textwidth]{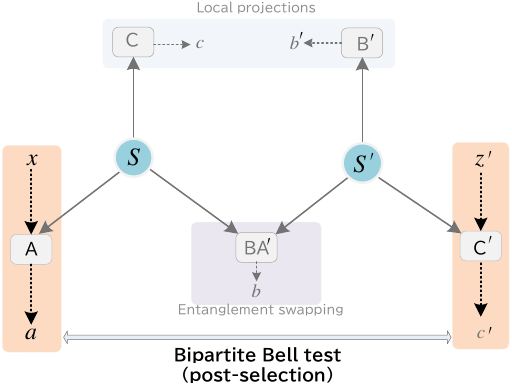}
\end{center}
\caption{Schematical Bell test based on entanglement swapping on a non-symmetric 5-partite network. One source $S$ distributes states to parties $\sA,\sB$ and $\sC$, respectively, while an identical copy $S^{\prime}$ distributes the same states to parties $\sA', \sB'$ and $\sC'$, respectively. Parties $\sC$ and $\sB'$ (grey box) perform local projections with respective outcomes $c$ and $b'$. Parties $\sB$ and $\sA'$ (purple box) are connected to perform to a joint measurement with outcome $b$, i.e., implementing entanglement swapping in the quantum realization. Parties $\sA$ and $\sC'$ (orange box) perform a standard Bell test on the post-selected bipartite entangled state with outcomes $a$ and $c'$ depending on measurement settings $x$ and $z'$, respectively.}
\label{fig00}
\end{figure}

In this Letter, we propose a network-based approach to generalize Gisin's theorem under the assumption of multiple copies, proving all genuine multipartite entangled pure states are GMN within Svetlichny’s biseparable non-signalling (NS) model \cite{Sy}. Our strategy involves a Bell test on a specific inflated network consisting of $n-1$ copies of the state under investigation, with each part performing one or two measurements. We show that the quantum correlations in this network can maximally violate a generalized CHSH inequality \cite{CHSH}, which holds for any realization of the network using biseparable NS sources \cite{Sy}. We further develop an alternative method that uses only two copies of the source but requires more than two measurement settings per party. In contrast to recent methods \cite{Contreras2021,Contreras2022}, both of our methods are noise-robust and capable of verifying the GMN of a single-copy state. Experimentally, we construct a hybrid photonic quantum network that encodes qubits in both polarization and path degrees of freedom (DoFs), enabling the preparation and measurement of an inflated network built from two copies of three-qubit states. Using this platform, we verify genuine tripartite nonlocality (GTN) for both GHZ-class and W-class states \cite{Dur} within previously inaccessible parameter regimes \cite{Ghose2009,Zhang2016,Wang2016,Wang2025}, significantly extending the scope of experimental verification for multipartite quantum nonlocality.

\textbf{Theories}. The standard Bell test for two or multiple parties involves a source distributing states to spacelike-separated observers $\sA_1, \cdots, \sA_n$. The measurement outcome $a_i$ of $\sA_i$ depends on local shared hidden variables and the measurement setting $x_i$. A joint probability distribution of all outcomes, conditional on measurement settings, is considered GMN if it cannot be decomposed into a biseparable form \cite{Sy}:
\begin{eqnarray}
P_{bs}(\va|\vx)
=\sum_{I,\oI}p_{I,\oI}P(\va_I|\vx_I) P(\va_{\oI}|\vx_{\oI}),
\label{eqn-2}
\end{eqnarray}
where $\va=(a_1,\cdots, a_n)$, $\vx=(x_1,\cdots, x_n)$, $\va_{J}=(a_i,i\in J)$,  $\vx_{J}=(x_i,i\in J)$, $I$ and $\oI$ denote a bipartition of $\{1, \cdots, n\}$, and $\{p_{I,\oI}\}$ is a probability distribution over all bipartitions. Here, $P(\va_I|\vx_I)$ is the joint distribution of outcomes $\va_I$ given measurement settings $\vx_I$, and similarly for $P(\va_{\oI}|\vx_{\oI})$. Both correlation sets $\{P(\va_I|\vx_I)\}$ and $\{P(\va_{\oI}|\vx_{\oI})\}$ satisfy the NS principle \cite{PR,Gal,Ban}, which means that $\sum_{a_j}p(a_i, a_j|x_i,x_j)=\sum_{a_{j'}}p(a_i, a_{j'}|x_i,x_{j'})$ for any $i$, $j$ and $j'$. 

A multipartite state is GME if it cannot be decomposed into a mixture of biseparable states \cite{Sy}:
\begin{eqnarray}
\rho_{bs}=\sum_{I,\oI}p_{I,\oI}\rho_{I,\oI},
\label{eqn-2a}
\end{eqnarray}
where $\rho_{I,\oI}$ is separable with respect to the bipartition $I|\oI$ of the Hilbert space $\cH_I\otimes \cH_{\oI}$, $\cH_I =\otimes_{i\in I}\cH_i$ and $\cH_{\oI} =\otimes_{j\in \oI}\cH_j$.

To illustrate our main idea, we first make the following source assumption.

\textbf{Assumption 1}. In each round of the Bell experiment, two identical copies of a given multipartite state are available and distributed to independent nodes of the network.
 
\textit{Tripartite systems.} Under Assumption 1, we consider the inflated network shown in Fig.~\ref{fig00}. Different from methods that characterizes the joint distribution with one copy within an inflated causal network \cite{causal2019}, we examine the joint distribution of two parties in different copies. This setup can be regarded as a Bell test in an entanglement swapping configuration \cite{ES,pan1998experimental,BGP}. It utilizes two identical sources, $S$ and $S'$, each distributing a  tripartite state to observers $\{\sA, \sB, \sC\}$, $\{\sA'$, $\sB'$, $\sC'\}$, respectively. Observers $\sB$ and $\sA'$ are connected to perform a joint operation (denoted as $\sB$), effectively constituting a Bell test on a 5-partite network. 

We first characterize the network in Fig.~\ref{fig00} realized with two identical copies of biseparable NS sources. Within the NS framework \cite{PR,BLM}, any biseparable NS source is represented by a mixture of separable sources (see Supplemental Material (SM) A.1 \cite{SI}): $\dt_{ABC}=p_1\dt_A\bot\dt_{BC}+ p_2\dt_B\bot\dt_{AC}+ p_3\dt_{AB}\bot\dt_C$, where $\dt_{AB(BC/AC)}$ are bipartite NS sources and $\dt_{A(B/C)}$ denote local sources, and $\{p_1,p_2,p_3\}$ is a probability distribution. Similar to classical and quantum theories, this source can be used to perform a Bell-type experiment and generate correlations. Specifically, suppose observers $\sC$, $\sB$, and $\sB'$ each perform measurements ${C^c}$, ${B^{b}}$, ${\hB^{b'}}$ with outcome $c, b, \hb \in \{0,1\}$, respectively, where $\sum_cC^c=\sum_{b} B^{b}=\sum_{b'}\hB^{b'}=\mathbbm{1}$ with the identity operator $\mathbbm{1}$. The other two observers $\sA$ and $\hsC$ {perform} local measurements $\{M^a_x\}$ and $\{M^{c'}_{{z^{\prime}}}\}$ with outcomes $a,c'\in \{0,1\}$, respectively, satisfying $\sum_aM^a_x=\mathbbm{1}$ and $\sum_{c'} M^{c'}_{{z^{\prime}}}=\mathbbm{1}$ for any $x$ and ${z^{\prime}}$. We prove that the resulting joint probability $P(a,b,c,b',c'|x,{z^{\prime}})$ of all outcomes, conditioned on the measurement settings of $\sA$ and $\sC'$, satisfies the following Bell inequality (see SM A.2 \cite{SI}). 

\textbf{Lemma 1}. For any outcomes $b, c$ and $b'$, the joint probability distribution generated by the Bell test on the network in Fig.~\ref{fig00} with two identical copies of biseparable NS sources satisfies:
\begin{eqnarray}
	\mathcal{L}_3&\equiv &\sum_{x,{z^{\prime}}=0,1}(-1)^{x\cdot{}{z^{\prime}}}\langle A_x B^{b} C^{c} B'^{b'}  C'_{{z^{\prime}}} \rangle\nonumber\\
	&\leq &\mathcal{R}_3\equiv 2.5 \langle B^{b} C^{c} B'^{b'}\rangle,
	\label{eqn-6}
\end{eqnarray}
where the correlators are defined by $\langle A_xB^{b} C^{c} B'^{b'} C'_{{z^{\prime}}}\rangle=\sum_{a,{c^{\prime}}=0,1}(-1)^{a+{c^{\prime}}}P(a,b,c,b',c'|x,{z^{\prime}})$ based on the joint distributions $\{P(a,b, c,b',c'|x,{z^{\prime}})\}$. The quantum observables are defined by $A_x=M_x^{0}-M_x^{1}$ and $C'_{{z^{\prime}}}=M_{{z^{\prime}}}^{0}-M_{{z^{\prime}}}^{1}$.

\begin{figure*}[ht!]
	\centering
	\includegraphics[width=17.5cm]{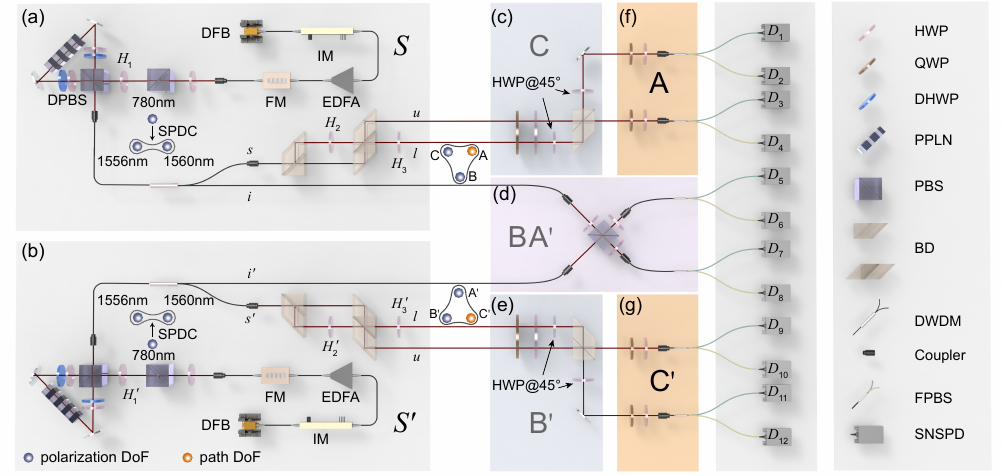}
	\caption{Schematic of the experimental setup. (a) and (b) show the setups for preparing two copies of generalized GHZ or W states, $S$ and $S'$. (d) shows the setup for implementing the joint measurement on qubits \textsf{B} and ${\textsf{A}}^{\prime}$ encoded on two idler photons. (c) and (e) show setups for measuring the polarization qubits \textsf{C} and ${\textsf{B}}^{\prime}$. (f) and (g) show setups for measuring the path qubits \textsf{A} and ${\textsf{C}}^{\prime}$. DFB, Distributed Feedback Laser; IM, Intensity Modulator; EDFA, Erbium-Doped Fiber Amplifier; FM, Frequency Multiplier; DPBS, dual-wavelength PBS; DHWP, dual-wavelength HWP.}
	\label{ex-set}
\end{figure*}

The key of the proof is to show the bipartite source shared between parties $\sA$ and $\sC'$ conditional on any outcomes of the other parties is a mixture of a bipartite NS source \cite{PR,BLM} and a separable source, i.e., $\dt_{AC'}=p\dt_{AC'}+ (1-p)\dt_A\bot\dt_{C'}$ with $p\leq 1/4$. The result then follows by combining this decomposition with the classical and NS bounds of the CHSH correlator \cite{CHSH}. See SM A.2 for details \cite{SI}.

Lemma 1 provides a direct method for verifying GTN. Specifically, for any genuinely entangled tripartite pure state $\ket{\Phi}$ on Hilbert space $\cH_A\otimes \cH_B\otimes \cH_c$, by employing generalized entanglement swapping \cite{ES}, observers $\sA$ and $\sC'$ can post-select a maximally entangled EPR state with assistance of local projection measurements of the other observers. Combining this with the CHSH test \cite{CHSH} implies that for any genuinely entangled state $\ket{\Phi}$ there are local observables for all parties such that the generated quantum correlations satisfy 
$\mathcal{L}_3 =2\sqrt{2}\langle B^{b} C^{c} B'^{b'}\rangle>\mathcal{R}_3=2.5\langle B^{b} C^{c} B'^{b'}\rangle$, which maximally violates the inequality (\ref{eqn-6}). 

\begin{theorem}
\label{Theorem1}
Any tripartite pure state that is genuinely multipartite entangled is genuine multipartite nonlocal under Assumption 1.
\end{theorem}

The proof of Theorem~\ref{Theorem1} is not restricted to the specific network configuration shown. For example, a similar result holds if party $\sA$ , receiving a state from the first copy, and party $\hsC$ , receiving a state from the second copy, perform a joint measurement. This means our scheme provides an efficient device-independent method for verifying the genuine tripartite nonlocality of all genuinely entangled pure states under Assumption 1. 

\textit{Multipartite systems.} Theorem 1 can be further generalized to any multipartite pure states under the following extended assumption.

\textbf{Assumption 2}. In each round of the Bell experiment, $n-1$ identical copies of a given $n$-partite state are available and distributed to independent nodes of the network.

Under Assumption 2, we consider an extended Bell-type test on a generalized chain network comprising $n-1$ copies of the given source $S$, as shown in SM B.1 \cite{SI}. For any genuine $n$-partite entangled pure states, assisted by local projections of specific nodes, a maximally entangled EPR state can be post-selected between the two end nodes of the terminal copies. This EPR state can then be used to achieve a maximal violation of the CHSH inequality \cite{CHSH}. However, the correlations generated by biseparable NS sources in this network cannot achieve this maximal violation (see SM B.1 \cite{SI}). An alternative method uses Assumption 1 and extends the chain Bell inequality \cite{Braunstein}, at the cost of only two copies but more than two measurements per party, see SM B.2 \cite{SI}.

\begin{theorem}
	\label{Theorem2}
   The GME of any multipartite pure state is GMN: (i) under Assumption 2 with no more than two measurements per party, or (ii) under Assumption 1 with more than two measurements per party.

\end{theorem}

Theorem 2 presents a trade-off method in GMN verification for GME states. The first method, utilizing $n-1$ copies, requires no more than two measurement settings per party but becomes resource-intensive for large $n$. The second method, using only two copies, is more experimentally feasible but requires an increased number of measurement settings per party and offers reduced robustness to noise compared to the first approach.

\textbf{Experiments}. We experimentally construct a hybrid photonic quantum network to test the GTN of the generalized GHZ state, $\ket{\textsf{GGHZ}}=\cos\theta\ket{000}+\sin\theta\ket{111}$ with $\theta\in (0,\frac{\pi}{2})$ and generalized W state, $\ket{\textsf{GW}}=a\ket{001}+b\ket{010}+c\ket{100}$ with $a^2+b^2+c^2=1$. The qubits are encoded in the polarization and path DoFs of photons.

The experimental setup is depicted in Fig.~\ref{ex-set}. We employ a Sagnac interferometer with a type-0 periodically poled MgO-doped lithium niobate (PPLN) crystal to generate two-photon state $\cos \theta |H_sH_i\rangle +\sin \theta |V_sV_i\rangle$. Then, we direct the signal photons through a beam displacer (BD) to introduce the path DoF, realizing two copies of generalized 3-qubit GHZ states:
\begin{equation}
	\label{GHZ-class1}
	\begin{aligned}
		\ket{\textsf{GGHZ}}&=(\cos \theta \ket{u_sH_iH_s} +\sin \theta \ket{l_sV_iV_s})_{{{\textsf{A}}}{\textsf{B}}{\textsf{C}}},
\\
\ket{\textsf{GGHZ}'} &=(\cos \theta \ket{H_{i'}H_{s'}u_{s'}} +\sin \theta \ket{V_{i'}V_{s'}l_{s'}})_{{{\textsf{A}^{\prime}}}{\textsf{B}^{\prime}}{\textsf{C}^{\prime}}},	\end{aligned}
\end{equation}
where $0<\theta<\frac{\pi}{2}$, as shown in Fig.~\ref{ex-set}(a,b). Next, we perform local and joint measurements on the qubits of these two copies as shown in Figs.~\ref{ex-set}(c-g). Experimental details are provided in the End Matter and SM D \cite{SI}.

\begin{figure}[!htbp]
	\centering
	\includegraphics[width=\linewidth]{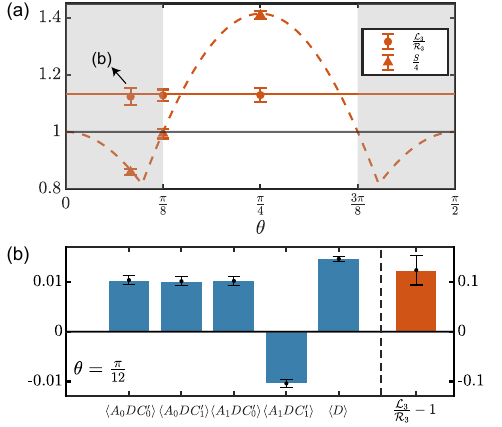}
	\caption{Experimental results for testing GHZ-class states. (a) The values of the ratio (red circles) between experimentally measured $\mathcal{L}_3$ and $\mathcal{R}_3$ for the generalized GHZ states with  $\theta \in \left\{ \frac{\pi}{12}, \frac{\pi}{8},\frac{\pi}{4} \right\}$. Red triangles show the values of the Svetlichny test $\frac{\mathcal{S}}{4}$ yielded with optimal measurements. The red solid line and dashed line represent corresponding ideal results obtained by numerical simulation. The gray region represents the failure regime of the Svetlichny test for tripartite generalized GHZ states. (b) The values of measured correlators and the violations of inequality (\ref{eqn-6}) with generalized GHZ states with $\theta =\frac{\pi}{12}$, $\frac{\mathcal{L}_3}{\mathcal{R}_3}-1$, where $ D = B^{b} C^{c} B'^{b'}$.  
	}
	\label{F1-GHZ}
\end{figure}

\begin{figure}[!htbp]
	\centering
	\includegraphics[width=\linewidth]{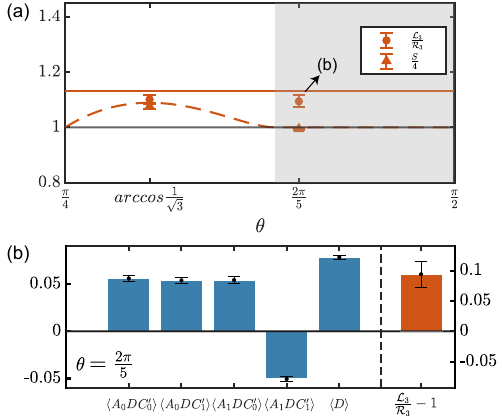}
	\caption{Experimental results for testing W-class states. (a) The ratio (red circles) between experimentally measured $\mathcal{L}_3$ and $\mathcal{R}_3$ for the generalized W states with $\theta \in\{ \mathrm{arccos} \frac{1}{\sqrt{3}}, \frac{2\pi}{5}\}$. Red triangles show the values of the Svetlichny test $\frac{\mathcal{S}}{4}$ yielded with optimal measurements. The red solid line and dashed line represent corresponding ideal results obtained by numerical simulation. The gray region represents the failure regime of the Svetlichny test for tripartite generalized W states. (b) The values of measured correlators and the violations of inequality (\ref{eqn-6}) with generalized W states with $\theta =\frac{2\pi}{5}$, $\frac{\mathcal{L}_3}{\mathcal{R}_3}-1$, where $D =B^{b} C^{c} B'^{b'}$.  
	}
	\label{F2-W}
\end{figure}

In the experiment, for each measurement setting, we collected data over a $1.5$-hour period, obtaining a series of probability distributions $\{P(a,b, c,\hb,\hc|x,\hz)\}$ to calculate $\mathcal{L}_3$ and $\mathcal{R}_3$. In Fig.~\ref{F1-GHZ}, we present the ratio between experimentally measured $\mathcal{L}_3$ and $\mathcal{R}_3$ for all three generalized GHZ states with $\theta \in \left\{ \frac{\pi}{12}, \frac{\pi}{8},\frac{\pi}{4} \right\}$ (Fig.~\ref{F1-GHZ}(a)), and the corresponding values of all correlators (Fig.~\ref{F1-GHZ}(b)). Results indicate that all these three generalized GHZ states exhibit genuine tripartite nonlocalities with negligible experimental noise. To further compare our method with previous methods, we test the Svetlichny inequality in the single-copy scenario. The results reveal that only the state with $\theta = \frac{\pi}{4}$ violates the inequality, while the states with $\theta = \frac{\pi}{8}$ and $\frac{\pi}{12}$ yield $\frac{\mathcal{S}}{4} \leq 1$ even with the optimal measurement strategy, indicating that the Svetlichny inequality cannot confirm genuine tripartite nonlocalities in these cases (See SM E and F \cite{SI}).

Moreover, by adjusting the angles of certain HWPs and adding an HWP@45$^\circ$ before the photons $i$ and $i'$, we simultaneously prepare two copies of generalized W states:
\begin{equation}\label{W-class}
	\begin{aligned}
		\ket{\textsf{GW}}=&(\cos \theta |{H_{s}}{H_{i}}l_{s}\rangle+\cos \theta |{H_{s}}{V_{i}}u_{s}\rangle
		\\
		&+\sqrt{-2\cos 2\theta}|{V_{s}}{H_{i}}u_{s}\rangle) _{\textsf{ABC}},
		\\
		\ket{\textsf{GW}'}=&(\cos \theta |{H_{i'}}{H_{s'}}l_{s'}\rangle+\cos \theta |{H_{i'}}{V_{s'}}u_{s'}\rangle 
		\\
		&+\sqrt{-2\cos 2\theta}|{V_{i'}}{H_{s'}}u_{s'}\rangle)_{{{\textsf{A}^{\prime}}}{\textsf{B}^{\prime}}{\textsf{C}^{\prime}}},
	\end{aligned}
\end{equation}
where $\frac{\pi}{4}<\theta<\frac{\pi}{2}$. Here, we allow the first two parameters of the state $\ket{\textsf{GW}}$ to change synchronously without loss of generality. The results are shown in Fig.~\ref{F2-W}, and indicate the general applicability of our method for various genuine multipartite entangled states, including both GHZ-class and W-class states (See SM D \cite{SI}).

Furthermore, we demonstrate that our method is significantly robust against various experimental noises, including white noise and decoherence noise (see SM C \cite{SI}). Experimentally, we introduce decoherence noise by inserting fused silica plates (see SM D \cite{SI}). The results show that our method maintains consistent robustness for generalized GHZ and W states under decoherence noise, respectively (see SM E \cite{SI}). To quantify the similarity of two copies, we evaluate the relative fidelity between them, all exceeding $0.975$ (see SM G \cite{SI}).

\textbf{Discussions}. The verification of GMN of multipartite entanglement depends on specific Bell tests. In this work, we introduced a novel approach  based on an extended Bell experiment performed on a network inflated with multiple identical copies of a given source. We highlight that although our method relies on the multi-copy assumption similar to previous results \cite{Contreras2021,Contreras2022}, it certifies the GMN inherent in single copies. Moreover, the present inequality for multipartite case may be further improved by combining different inflation methods (see SM H \cite{SI}). Furthermore, since genuine multipartite steering (GMS) \cite{HR,cavalcanti2015detection,uola2020quantum} exhibits activating effects on Bell tests, it represents a form of correlations weaker than GMN but may be stronger than GME. Our findings thus demonstrate the equivalence of three types of correlations of GME, GMS, and GMN for any isolated system under the multi-copy source assumption (see SM I \cite{SI}). Another related problem is to consider the GMN under alternative physical models \cite{coiteux2021no, coiteux2021any, PhysRevLett.129.150401, cao2022experimental}.

In our experiment, using inequality (\ref{eqn-6}), we accomplished the first experimental verification of GTN across a wide range of GHZ-class and W-class states, different from previous experiments \cite{Zhang2016,Wang2016}. Moreover, we demonstrate that our method exhibits strong robustness against experimental noise, achieving identical noise visibility for generalized GHZ and W states under decoherence. Our experiments demonstrate the feasibility and advantages of GMN verification within multi-copy inflated networks. A key assumption is the indistinguishability of the sources, as all are modeled under the same Svetlichny biseparable framework. Moreover, we emphasize that additional loopholes remain unaddressed. Specifically, the locality loophole persists due to two qubits being encoded in the same photon, and the experiment is also subject to the freedom-of-choice and detection loopholes \cite{RMP}. Our work not only represents a significant advancement in quantum foundations by introducing new avenues for exploring GMN, but also deepens the understanding of quantum correlations in multipartite systems and networks.

\begin{acknowledgments}
\textit{Acknowledgments---}This work was supported by the Quantum Science and Technology-National Science and Technology Major Project (No.~2023ZD0301200), the National Natural Science Foundation of China (Nos.~T2522017, 12574394, 92476103, 62375117, 62172341, 12204386, 12405024, 12075159), the Guangdong Basic and Applied Basic Research Foundation (No.~2025B1515020064), Sichuan Natural Science Foundation (Nos.~2024NSFSC1375, 2024NSFSC1365), the Tianjin Natural Science Foundation Project (No.~25JCQNJC01170), the Shenzhen Science and Technology Program (No.~KQTD20200820113010023) and the Shenzhen Fundamental Research Program (No.~JCYJ20220530113404009), Interdisciplinary Research of Southwest Jiaotong University China (No.~2682022KJ004), and the Academician Innovation Platform of Hainan Province.

\textit{Data Availability---}The data that support the findings of this article are openly available \cite{GTN_data}.
\end{acknowledgments}

\clearpage 

\newpage

\section{End Matter}

The experimental setup is depicted in Fig.~\ref{ex-set}. Beginning with the source S shown in Fig.~\ref{ex-set}(a), we illustrate the preparation process of the first copy of the generalized GHZ state. Firstly, we generate a series of laser pulses with a pulse duration of 66 ps and a central wavelength of 779 nm using a laboratory-assembled laser module. This module comprises a distributed feedback (DFB) laser and an intensity modulator (IM) driven by a pulse pattern generator, followed by an erbium-doped fiber amplifier (EDFA) and a frequency multiplier (FM). By configuring the half-wave plate (HWP) $H_1$ at an angle $\alpha=\frac{\theta}{2}$, we can prepare the photons of each pulse into a classical electric field superposition state $ \cos\theta E_H + \sin\theta E_V$, where $E_H$ and $E_V$ represent the classical electric field components for horizontal and vertical polarizations, respectively. Then the photons of each pulse in state $|H\rangle$ and $|V\rangle$ bidirectionally pass through a periodically poled MgO-doped lithium niobate (PPLN) crystal placed in a Sagnac interferometer, respectively, to generate two polarization-entangled photons via type-0 spontaneous parametric down-conversion (SPDC) processing \cite{sun2019experimental, PhysRevLett.128.040402}. The central wavelengths of the signal and idler photons are 1556 nm and 1560 nm, respectively, enabling them to be separated by a dense wavelength division multiplexing (DWDM) filter. The two-photon polarization state can be characterized as $\cos \theta |H_sH_i\rangle +\sin \theta |V_sV_i\rangle$. Next, we direct the signal photons through a beam displacer (BD) to introduce the path DoF. The BD is composed of a polarization beam splitter (PBS) and a mirror, which transmits photons in the $|H\rangle$ state and reflects those in the $|V\rangle$ state, respectively. Then, the state can be characterized as $\cos \theta |H_sH_il_s\rangle +\sin \theta |V_sV_iu_s\rangle$, with $\ket{u_s}$ and $\ket{l_s}$ denoting the upper and lower path of the signal photons as shown in Fig.~\ref{ex-set}(a). After that, we introduce a second BD consisting of a polarization beam splitter (PBS) and two mirrors sandwiched between two HWPs $H_2$ and $H_3$, oriented at angles $\beta=\frac{\pi}{4}$ and $\gamma=\frac{\pi}{4}$, respectively. This configuration enables the swapping of photon paths based on their polarization states. The state is finally transformed to a generalized 3-bit GHZ state, 
\begin{equation}
	\label{GHZ-class1}
	\begin{aligned}
		\ket{\textsf{GGHZ}}=(\cos \theta \ket{u_sH_iH_s} +\sin \theta \ket{l_sV_iV_s})_{{{\textsf{A}}}{\textsf{B}}{\textsf{C}}}, 
	\end{aligned}
\end{equation}
encoded in the polarization and path DoFs of two photons, then we complete the preparation process.
Here, corresponding to the network of Fig.~\ref{fig00}, the qubits \textsf{A} and \textsf{C} are encoded in the path and polarization DoFs of the signal photon $s$, while the qubit \textsf{B} is encoded in the polarization DoF of idler photon $i$.

Similarly in Fig.~\ref{ex-set}(b), we prepare the second copy of the generalized 3-qubit GHZ state 
\begin{equation}\label{GHZ-class2}
	\begin{aligned}
		\ket{\textsf{GGHZ}'} =(\cos \theta \ket{H_{i'}H_{s'}u_{s'}} +\sin \theta \ket{V_{i'}V_{s'}l_{s'}})_{{{\textsf{A}^{\prime}}}{\textsf{B}^{\prime}}{\textsf{C}^{\prime}}}.
	\end{aligned}
\end{equation}
by setting the three adjustable HWP $H'_1$, $H'_2$ and $H'_3$ at angles $\alpha'=\frac{\theta}{2}$, $\beta'=\frac{\pi}{8}$ and $\gamma'=\frac{\pi}{8}$, respectively. Here the qubit \textsf{A} is encoded in the polarization DoF of idler photon $i'$, the qubits \textsf{B} and \textsf{C} are encoded in the polarization and path DoFs of the signal photon $s'$, respectively (see SM D and G \cite{SI}).

 Next, we perform the local and joint measurements on the qubits of these two copies. As shown in Fig.~\ref{ex-set}(d), after an HWP$@45^\circ$ flipping the polarization of photon $i$, we let the photons $i$ and $i'$ overlap at a PBS and measure them in each output port using a combination of an HWP$@22.5^\circ$ and a fiber PBS (FPBS), followed by two superconducting nanowire single-photon detectors (SNSPDs), then we complete a partial Bell state measurement \cite{pan1998experimental} on the qubits \textsf{B} and ${\textsf{A}}^{\prime}$, and project the photons in states $|\varPsi ^+\rangle$ with the coincidence events between detectors $D_5 \& D_7$ or $D_6 \& D_8$. Additionally, as shown in Figs.~\ref{ex-set}(c) and \ref{ex-set}(e), we project each of the polarization qubits \textsf{C} and ${\textsf{B}}^{\prime}$ of photons $s$ and $s'$ into state $\ket{+}$ using an QWP\&HWP combination acting on both paths, with an additional HWP$@45^\circ$ installed solely on the lower path before the BD. The photons outcome from the two ports of the BD,  indicates they are projected into states $\ket{\pm}$, respectively, where $\ket{\pm}=\frac{\ket{0}\pm\ket{1}}{\sqrt{2}}$, and after that the qubits \textsf{A} and ${\textsf{C}}^{\prime}$ encoded in path DoF of the signal photons are converted into the polarization DoF through above operations. Finally, we measure the qubits \textsf{A} and ${\textsf{C}}^{\prime}$ with the observer $A_x\in \left\{ \sigma _y, \sigma _x \right\}$ and $C'_{z'}\in \left\{ \frac{\sigma _y+\sigma _x}{\sqrt{2}},\frac{\sigma _y-\sigma _x}{\sqrt{2}} \right\}$, $\sigma _x$ and $\sigma _y$ are Pauli matrices. As shown in Figs.~\ref{ex-set}(f) and \ref{ex-set}(g), the measurement device includes an adjustable HWP-QWP pair, followed by an FPBS for projective measurements.
\end{document}